\documentclass[aps,nofootinbib,twocolumn,citeautoscript]{revtex4}
\usepackage{amsmath,amssymb}
\usepackage{amsmath,amssymb}
\usepackage{enumerate}
\usepackage{graphicx}
\usepackage{xcolor}
\usepackage{enumerate}
\usepackage{gensymb}
\usepackage[colorlinks=true,citecolor=red,linkcolor=blue,linkbordercolor=cyan]{hyperref}

\usepackage{float}
\date{\today}
\usepackage{float}
\date{\today}
\begin{document}

\title{Quantum Machine Learning and its Supremacy in High Energy Physics}

\author{Kapil K. Sharma$^{\dagger}$\\
\textit{DY Patil International University,\\
Sect-29, Nigdi Pradhikaran\\
Akurdi, Pune,\\
Maharashtra-411044, India} \\
E-mail: iitbkapil@gmail.com$^{\dagger}$}

\begin{abstract}
This article reveals the future prospects of quantum algorithms in high energy physics (HEP). Particle identification, knowing their properties and characteristics is a challenging problem in experimental HEP. The key technique to solve these problems is pattern recognition, which is an important application of machine learning and unconditionally used for HEP problems. To execute pattern recognition task for track and vertex reconstruction, the particle physics community vastly use statistical machine learning methods. These methods vary from detector to detector geometry and magnetic filed used in the experiment. Here in the present introductory article, we deliver the future possibilities for the lucid application of quantum computation and quantum machine learning in HEP, rather than focusing on deep mathematical structures of techniques arise in this domain.
\end{abstract}
\maketitle
\section{Introduction}
The field of high energy physics (HEP) deals with the discovery of varieties of particles which gives the clue to understand the big bang and origin of universe\cite{hep1,hep2}. The HEP experiments\cite{ehep} demand the high voltage to operate and need the accelerators for beam collisions. At CERN\cite{cern}, the Large Hadron Collider (LHC) is the biggest particle collider in the world, which has been operated with energy (6.5 TeV/beam) in its second run scheduled in 2015. However a short run of the accelerator with  xenon-xenon collisions has been performed in 2017. The experimental 
set-up of LHC is tunnelled underground at 175 meters, it has its huge diameter as 27 kilometres. In LHC the particle beams are launched in anti directions at very high speed, the beams further collide at many interaction points available at the periphery of large accelerator. The interaction points support the bombardment of antiparticle beams. This bombardment release huge energy with different kind of particles and varieties of trajectories are developed by particles during this process. The most important ingredient used to capture the event happened at interaction point is the detector. Round the periphery of the accelerator, there are seven detectors (ATLAS, CMS, LHcb, ALICE, TOTEM, LHcf, MoEDAL) assembled, which are used for different roles in LHC\cite{lhc1,lhc2}. Experimental part of HEP involve many complexities in terms of designing detectors, high-end electronics, data acquisition systems and software\cite{data}. The data gather in real time at LHC is recorded at a tape and processed through grid computing, which further can be distributed to many universities and research centres for particle physics analysis. The changes in designing methodologies and implementation of various detectors is very crucial and important part. The detectors play an important role to capture the event and provide the huge data corresponding to interaction points. The technical journey of detectors from bubble chamber to semiconductor detectors have long strides. Each detector has association with front-end electronics equipped with data acquisition system (DAS)\cite{data}. The DAS gather the information from detector in real time during the event happening in accelerator. It also play the role to avoid the unnecessary background events and to collect only valid events during the triggering process. For this purpose the triggering may be implemented at many levels of hardware in real time. Most of the time DAS suffer from dead time, the time during which no event is captured and there is also the possibilities of missing the events. The dead time in DAS depends on many factors such as clock speed of the electronics circuitry, noise, rate of event happening etc. Because of the dead time, the speed of writing the data on storage tape is also affected. Once the data is recorded on tape through grid computing, it can be
distributed further to do offline analysis to extract the information about particle trajectories developed inside the detectors\cite{data}. These trajectories are important ingredient which have the hidden information about many characteristics of the particles. During the analysis of offline data the machine learning come into picture and play the important role\cite{mch1,mch2}. The point where initially two anti beams collide is called primary vertex, while secondary vertex may also be produced because of the particle decay inside the LHC\cite{track1}. The tracks produced inside the detectors can have complex structure. In particular, the complex structure of tracks arise because of the
magnetic field associated with LHC solenoid. In this process of colliding of two anti directional beams inside the LHC, there are always chances for high background noise for which the material of the detector has the significant impact. The process to determine the particle characteristics depends on track reconstruction and their fitting, this process is governed by pattern recognition methods performed on offline data and no doubt machine learning techniques contribute a lot. There are important software, which perform these tasks such as GEANT4, ROOT, HERWIG etc\cite{soft1,soft2,soft3}. These software use the classical algorithms, which of course are detector dependent and take many features of detector into account such as it's geometry, orientation, diameter, its material etc. Here it is mentioned that machine learning played an important role to solve the problem of track reconstruction and fitting in HEP from a decade\cite{ml1}. To the date, the techniques of machine learning (supervised, unsupervised) have been implemented in offline data
\begin{center}
\begin{figure}
\includegraphics[scale=0.3]{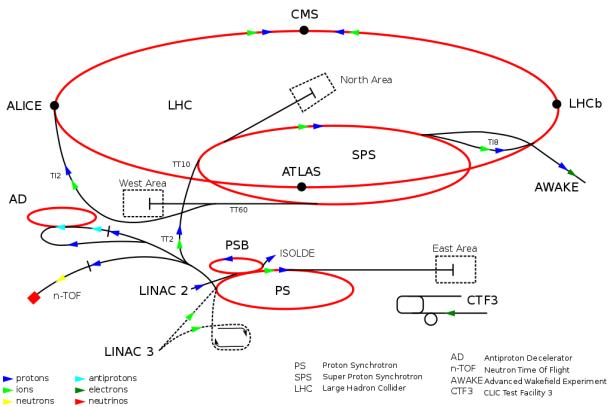}
\caption{The detectors at CERN.}
\end{figure}
\end{center}
simulation in HEP with many models such as neural network, deep learning, simulated annealing etc\cite{nl1,mch2,sm1}. These technique are successfully performed well to discovered the particle Higgs Boson\cite{higgs}, which is a great example. As high end electronics and fast algorithms is the primary requirement for HEP experiments. So, to overcome these issues, one can think towards HEP on quantum computer. Off course quantum hardware is not mature till date but there is a future hope for quantum processors\cite{dw1,dw2,dw3,dw4,dw5}. This may help to execute algorithms with significant speed and can lead the scenario towards quantum machine learning algorithms development which can be  utilized in HEP\cite{qml1,qml2,qml3,qml4,qml5,qml6,qml7,qml8,qml9,qml10}. There are landmark quantum algorithms such as (Shor's, Love Grover) algorithms\cite{sho,love}. These algorithms give the clue to develop many other quantum algorithms in the domain of machine learning and optimization used for varieties of tasks. The domain of developing quantum algorithms and studying the quantum complexity\cite{qcom} open the newly emerging field of quantum machine learning. Recently, there is rapid progress in this filed, St Loyed et al. have been developed the quantum algorithm to solve the system of linear equations on the quantum computer\cite{le1}, Fernando et al. developed the semi-infinite programming algorithm which is a step towards quantum algorithms for optimization problems\cite{le2,le3}. D-Wave systems have been developed quantum annealing based processor, which is an indication for future solutions of optimization problems on quantum hardware\cite{dw1,dw2,dw3,dw4,dw5}. 
To the date, there are huge attempts to investigate the properties of quantum counterpart models such as quantum neural networks\cite{n1,n2,n3}, quantum deep learning\cite{qml5,dp1}, quantum Boltzmann machines, quantum annealing and others. The area of quantum machine learning can serve better for many tasks performed in offline data simulation in HEP and set-up a new domain of research.

The paper is sketched in three sections. In Sect. 2, we discuss the Track reconstruction and machine learning techniques. Sect 3, is devoted for the supremacy of quantum machine learning in HEP.
\section{Track reconstruction and pattern recognition}
In this section, we introduce the method for track reconstruction and also give the shadow on machine learning techniques used for the same. The track reconstruction is the important requirement in HEP which can be divided into two basic steps as 1) finding the track candidates 2) track fitting\cite{data,track1}. The primary requirement of track fitting is that it must be robust against the error-prone of track finding procedure, it must be fast and numerically stable. Overall, it is important to mention that the track reconstruction strictly depends on the type of detectors used in HEP experiments. Most of the previously used detectors such as bubble chamber, gaseous chamber etc. are completely obsolete and overtaken by semiconductor detectors.\cite{data} In practical applications semiconductor suffers from radiation released in the collision of particle beams, hence to overcome this phase, the research to develop the diamond detectors is very active\cite{dd1,dd2}. As an example, the inner detector used in ATLAS use semiconductor technology and has complicated geometry. The assembling and installation process of detectors often disturb their geometry over the pre-assumed geometry. This problem is called misalignment problem in detectors, which is a key element to produce the track candidates\cite{al1,al2,al3}. Getting the best track candidates, contribute to the goodness of algorithm for track reconstruction. Here it is mentioned that obtaining the track candidates can also be called as feature extraction and a primary step to reconstruct the track, in this process the classification is done of all hit points by particles in track detector. Each class set has all hit points for a single track and each class is called track candidate. It is important to state that these track candidates many times carry the noise, in other words, the background hit points. During the track finding process based on a particular track model, the pattern recognition plays a significant role, which is the part of machine learning. Before applying any machine learning techniques, it is always better to reduce the dimensionality of the data gathered in the experiment, such that, overall the outcome of the goal must not be affected in terms of better classification and error reduction.  There are countless algorithms for dimensionality reduction\cite{dr1,dr2,dr3} and these always can be challenged. Any dimensional reduction algorithm is suitable for one problem but may not fit for another one.  So data dimensionality reduction technique in machine learning is a highly challenging step and must be performed carefully if required because the adoption of any bad technique always lead towards wastage of efforts. In spite of focusing on these methods in detail, here we discuss the method of track finding, which can be divided into two categories as local and global methods. In detail for these topics, the reader may refer (Ref:\cite{data}). In continuation of the paper, we proceed the short introduction of these methods in terms of offline data analysis. The track finding is the crucial part of track reconstruction and first needs the track modeling. Track modeling takes into consideration the geometry of the detector, associated magnificent filed, noise, measurement errors etc. Few important mathematical approximations with circles, parabola and splines have been used for the same\cite{data}, these methods require the speed of the calculation and also need interpolation or extrapolation techniques for prediction\cite{data}. The next step after track modeling is the track finding, which definitely uses machine learning methods\cite{pt}. During the collision when particles hit the detectors layers and ionize the detector material than hitting corresponds to a kind of measurement of particle which is recorded by the sensor assembled in the detector. The set of measurements recorded by the detector help to find the track candidates and have the information about the track traced by the particle. There may exist any situation such that, missing the tracked candidate, or there is no track candidate and there may be track candidates which do not belong to any track. For the sake of clarity here we rewrite, track finding methods can be divided into two categories as local or global methods\cite{pt}. In local or sequential methods, the track is reconstructed sequentially by taking a seed. The seed is a portion of a track got from the measurements done by the detector during the collisions. Generally, two track modeling approaches have been used in local methods such as track-road and track-following. In track road method a hollow cylinder of a desirable diameter is considered around the trajectory, the points fallen inside the track road are considered for analysis by using the pattern recognition methods. Often, the track road methods are slower than track following methods. Track following methods are valid while the track candidates are easily identified by human senses. Further, in the global method, the track candidates are supplied to the algorithm at once to produce the tracks. The order of track candidates do not matter, but the execution speed of algorithm in this process is low in comparison to local  methods. The computation time is taken by the global method is proportional to the number of candidates. There are many classical approaches have been used for track finding such as Hough transform\cite{hg1}, Conformal mapping\cite{con1}, Kalman Filter\cite{kalman}, Neural networks, Deep learning\cite{con2} etc. Based on the above discussion we would like to emphasize that machine learning techniques are highly important for experimental particle physics, which can not be ignored. So can we think of better situations than existing techniques? Yes, hope so. In the next section we cover the future perspective of quantum machine learning techniques in HEP domain.
\section{Neural networks and Track reconstruction}
In this section we present classical adaptive method for track finding\cite{tr}. Adaptive method are more simpler than tradition methods because they do not involve the description of detector geometry. Almost all the adaptive methods use the model of neural network and their various forms. Many forms of neural networks such as hope filed network, Boltzmann machine have been used for the same purpose.
The basic ingredient of neural network is a neuron and its firing mechanism, which use an activation function. Let suppose the neural network is simulated over an Ising system with $N$ spins connected by synaptic strength $w_{ij}$. A spin may the spin up and spin down configuration, ie. $s_{i}\in \{+1,-1\}$. The dynamics of the neural network may be given by applying the local updating rule as follows,
\begin{equation}
s_{i}=sig(\sum_{j}T_{ij}s_{j})
\end{equation} 
The matrix $T_{ij}$ involve in the above equation has the information about synaptic strength of whole network. By taking into account the configuration of spins over the network and the matrix of synaptic strength, the energy function of the network can be written as follows, 
\begin{equation}
E(S)=-\frac{1}{2}\sum_{ij} T_{ij} s_{i}s_{j}
\end{equation}
The simple strategy to obtain the solution of a problem is to encode the problem in neural network and optimize the energy function. There are many ways to optimize the energy function, a common way is to setup the Boltzmann distribution function with introducing the parameter as temperature, it is given as below,
\begin{equation}
P(T)=\frac{e^{-\frac{E}{KT}}}{Tr[e^{-\frac{E}{KT}}]}
\end{equation}
Where $P(T)$ is the probability of occupying the highest energy state at a particular temperature. Properties of Boltzmann distribution are very helpful to develop varieties of algorithms for optimization with heuristic approaches. By following the basic frame work of neural network here we provide the track reconstruction encoding method in neural networks developed by Denby and Peterson independently. In this approach they encode the track segments in neurons with some assumptions. These assumptions are encoded in terms of constraints in energy function established over the specified neural network. The neural network based algorithms have input as a data produced by the detectors and the goal is to develop the tracks which have the following constraints,
a) The tracks must be smooth b) The tracks must not kink c). The tracks must not have bifurcation.
Debny and Peterson encoded a track segment formed between to points $(i,j)$ in a neuron named as $s_{ij}$. Let suppose another track segment connected between the points $(k,l)$ with the neuron $s_{k,l}$. If the condition $\{s_{ij}=+1,s_{kl}=+1\}$, satisfy than the connections are on, if the condition is followed as $\{s_{ij}=-1,s_{kl}=-1\}$, than the connections are off and no track segment is formed. After the track reconstruction problem in neural network the energy function can be written as below,
\begin{equation}
E_{ijkl}=-\frac{1}{2}\sum_{ijkl}(T_{ijkl}^{cost}+T_{ijkl}^{constraint}) s_{ij}s_{kl}\label{3}
\end{equation} 
Two track segments may have connections if the  condition $(j=k)$ is satisfied. The smoothness of the segments are encoded in the angle between two track segments ie $\theta$ with the following eq.,
\begin{equation}
T_{ijkl}^{cost}=\delta_{jk}\frac{cos^{m}\theta_{ijl}}{r_{ij}+r_{jl}}\label{1}
\end{equation}
Further the bifurcation encoding is done in the following equation,
\begin{equation}
T_{ijkl}^{constraint}=-\frac{\alpha}{2}[\delta_{ik}(1-\delta_{jl})-\delta_{jl}(1-\delta_{ik})]-\frac{\beta}{2}[text]\label{2}
\end{equation}  
with, \textit{text}$\longrightarrow$\text{``global inhibition"}\\ \\
Where $(\alpha,\beta)$ are Lagrange multipliers. By introducing the Eq.\ref{1} and \ref{2} in Eq.\ref{3}, the final energy function can be written. The final energy function can be optimized by using any suitable method with by applying some cuts in the parameters on number of neurons $(N)$, angle parameter $(\theta)$ and circle with certain radius which put the restrictions on the length of segments. 

\section{Supremacy of quantum machine learning in HEP}
Quantum pattern recognition\cite{qp1,qp2,qp3} is an important application of quantum machine learning. It is obvious that there is always research progress in HEP to develop fast and better algorithms. Can quantum pattern recognition techniques help in HEP to deal with massive amount of data on the quantum computer and extract the useful information from the data? Here we mention that there is recent progress on quantum algorithms in many domains like, algebraic domain (Hidden subgroup problems)\cite{mm1,mm2}, semidefinite programming\cite{le2}, linear differential equations\cite{le22,mm22}, finite element methods\cite{le23} and in pattern recognition\cite{mm3}. There is major developments for quantum algorithms with black box model and query complexity\cite{mm1}. Query complexity is the quantum equivalent of classical decision tree model. However the development of quantum algorithms with adiabatic quantum computation\cite{ad1} model is on slow progress, as this approach does not have any suitable complexity model to calculate the  quality of the algorithm, which is an open problem. But there are future possibilities for the same, which can contribute for better quantum algorithm designing. During the track reconstruction process the problem at many stages can be mapped to suitable optimization problems\cite{op1,op2} which may be solved further by any suitable method. The research to solve quadratic binary optimization problems subjected to constrained or unconstrained are on the way by using quantum annealing\cite{qd1,qd2}. The quantum annealing exhibit better signatures to handle the problems on the quantum computer. The quantum algorithms developed based on quantum strategy (superposition and entanglement\cite{en1,en2}) and taking into consideration the geometrical aspects of detectors may be useful in comparison to classical algorithms in terms of (time, space) complexity and speed. On the other hand, as per the literature survey, the algorithmic development in HEP is less pervasive towards the existence of entanglement during the collision process inside any detector. Can we also have such algorithms which can catch the phenomena of entanglement inside the detector if exists, which can help further to understand the true nature of particles? Pattern recognition in many forms involving the image processing had been the part of particle physics community for a long time\cite{im1}. There is literature, number of papers and Ph.D work in which the community has been solved the problems of particle physics by using image processing techniques\cite{im2}. It is not worth to mention that the emergence of quantum image processing\cite{qim1,qim2} can also contribute to better improvement for experimental particle physics. However the current trend deals with neural network and deep learning methods, which overcome to the difficulty of feature extraction\cite{f1} as involved in traditional theory of pattern recognition, but these techniques involve huge complex structures of networks which further make the optimization problems very difficult. It is mentioned that, the development of quantum algorithms and studying quantum complexities in particle physics domain is really challenging but do not seem impossible. The rapid progress in quantum domain boost the hope for future possibilities to solve particle physics problems on quantum computer. We hope the present article gives the sufficient indication to the HEP and quantum community to make the development towards the aforementioned directions, which is almost untouched.
\section{Optimization on analog quantum computer}
Recently quantum annealer\cite{qa1,qa2} has been developed by D-Wave system, which is a kind of analog quantum computer and suitable for many optimization problems. Quantum annealing is inspired by simulated annealing\cite{sim1}, the basic difference is that quantum annealing often use
magnetic filed as a controlling parameter, while simulated
annealing use temperature. Both the approaches can be
used to simulate spin glass systems, these systems further can be used to encode NP complete problems. NP complete problems are generally decision making problems, which produce the answers in yes and no. It is found that quantum annealing has better advantage in
terms of complexity reduction of NP complete problems\cite{np1}
in comparison to simulated annealing. On the other hand
quantum annealing can also be introduced as as restricted
version of adiabetic quantum computation\cite{adia} which do not
allow tunneling inside the physical system. The tunneling mechanism with the framework of quantum annealing has an extra advantage to reduce the computational complexity. The manufactured D-Wave quantum computer has a good attempt to solve a subset of optimization problems involved in mathematics more efficiently than classical computers. With the promising progress in designing the D-Wave computer, the fundamental results exhibts that quantum hardware has speedup capability
to implement few quantum algorithms for solving some
mathematical problems. The field of quantum computation is progressing rapidly. Regardless of the results
achieved by D-Wave and its hardware with quantum annealing, fully programmable quantum computer operate much differently than D-Wave but initial results for universal quantum computer seems very promising. The method of operation of these universal quantum computers needs a full architecture incorporating with many
components like storage medium, buses to transport the
signal, Quantum arithmetic and logic unit and quantum
softwares.
\section{Mapping the problem to QUBO}
\begin{figure}
\includegraphics[scale=0.3]{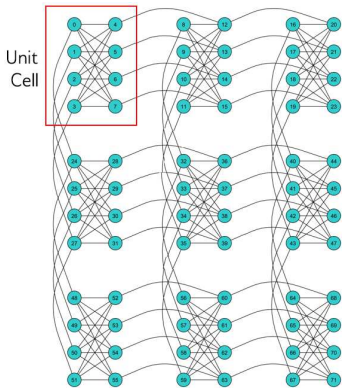}
\caption{Arrangement of qubits on Chimera graph with nine cells.}\label{f2}
\end{figure}

D-Wave quantum annealer have the connectivity of super-conducting qubits on Chimera graph. The graph has a collection of cells arranged in square fashion, each cell is consisted of by eight qubits connected in feed forward fashion. Further all the cells have special connections among each other. The arrangement of chimera graph is shown below in the Fig.\ref{f2}, 
Any quantum Ising systems can be mapped to quantum annealer for quadratic binary optimization (QUBO) problems. The method used for this mapping is called minor embedding\cite{me}. As per the definition of Ising system, it can be assumed as a set of spins in d-dimensional hyper cubic lattice. But for the current work, for simplicity we fix the limit over the
dimensions and assume that the spins are scattered in two dimensional
space at random in a variable magnetic filed over each spin. Every
spin has interaction with its neighbor spin modeled with certain coupling
strength. Please note that, here we are dealing with planer Ising
system, the scenario of non-planer Ising systems is different. The classical mathematical expression for planer Ising system in 2 dimensions can be written as below,
\begin{equation}
E(J|h)=\sum_{(i=1,j=1)}^{(N,N)}J_{ij}s_{i}s_{j}+\sum_{i}h_{i}s_{i}
\end{equation}
For quantum Ising system, one has to replace the variables $(s_{i},s_{j})$ with Pauli matrices as $(\sigma_{i},\sigma_{j})$ and the energy function $E(J|h)$ takes the from of an Hamiltonian operator. Here $J_{ij}$ is the pairwise coupling strength between $(i^{th},j^{th})$
spins and $(s_{i},s_{j})$ are the spin variables which can have eigenvalues
as $(\pm1)$. The quantity $(+1)$, denotes spin up and $(-1)$ denotes spin down. In Ising system, every spin can encode binary data, which is very useful to solve NP problems such as SAT problem, graph bipartition and many others. D-Wave quantum annealer use the adiabatic quantum computing strategy in which the system is prepared in initial ground state with the Hamiltonian $H_{I}$. The initial system is disturb by an additional Hamiltonian $H_{F}$. The total Hamiltonian of the system can be written as a convex combination as below,  
\begin{equation}
H(s)=(1-s)H_{I}+sH_{F}
\end{equation} 
The problem is encoded in $H_{F}$, whose ground state has the solution of the problem, provide the total  Hamiltonian $H(s)$ evolve with adiabatic condition\cite{ad}. Before adopting the quantum adiabatic evolutionary plan, it is advisable to encode the problem first on quadratic polynomial and further it can be easily mapped to available quantum annealer. Consider the quadratic polynomial over the three
spins $S=(s_{1},s_{2},s_{3})^{T}$ as,
\begin{equation}
(as_{1}+bs_{2}+cs_{3})^{2}=(as_{1}+bs_{2}+cs_{3}).(as_{1}+bs_{2}+cs_{3})
\end{equation}
\begin{equation}
=a^{2}(s_{1})^{2}+b^{2}(s_{2})^{2}+c^{2}(s_{3})^{2}+2ab(s_{1}s_{2})+2ac(s_{1}s_{3})+2bc(s_{2}s_{3})
\end{equation}
manipulating the Eq. (2) in matrix form we obtain,
\begin{eqnarray}
=[s_{1},s_{2},s_{3}].\begin{bmatrix}a\\
b\\
c
\end{bmatrix}.[a,b,c].\begin{bmatrix}s_{1}\\
s_{2}\\
s_{3}
\end{bmatrix}\\=[s_{1},s_{2},s_{3}].\begin{bmatrix}a^{2} & ab & ac\\
ab & b^{2} & bc\\
ac & bc & c^{2}
\end{bmatrix}.\begin{bmatrix}s_{1}\\
s_{2}\\
s_{3}
\end{bmatrix}=S^{T}QS
\end{eqnarray}
Here the matrix Q is symmetric, another representation of Q will also
give the same result as obtained in Eq. (3).
\begin{equation}
=S^{T}.\begin{bmatrix}a^{2} & 2ab & 2ac\\
0 & b^{2} & 2bc\\
0 & 0 & c^{2}
\end{bmatrix}.S
\end{equation}
Let decompose the matrix Q we get,
\begin{eqnarray}
=S^{T}.\{\begin{bmatrix}a^{2} & 0 & 0\\
0 & b^{2} & 0\\
0 & 0 & c^{2}
\end{bmatrix}+\begin{bmatrix}0 & 2ab & 2ac\\
0 & 0 & 2bc\\
0 & 0 & 0
\end{bmatrix}\}.S\\=S^{T}.(Q^{diag}+Q^{upper}).S\\=(S^{T}.Q^{diag}.S)+(S^{T}.Q^{upper}.S)
\end{eqnarray}
If the vector S is having N number of spins ie. $S\in\{s_{1},s_{2},s_{3}.......s_{N}\}$,
than the above expression can be written in generalized form as below,
\begin{eqnarray}
=\sum_{ii}S^{T}Q_{ii}^{diag}.S+\sum_{(i<j)}S^{T}.Q_{ij}^{upper}.S\\=\sum_{i\leq j}S^{T}.Q_{ij}.S
\end{eqnarray}
Proceeding for $N$ number of spins the quadratic form of a polynomial can be written as,
\begin{eqnarray}
(as_{1}+bs_{2}+cs_{3}+.........+\gamma s_{N})^{2} \nonumber \\=\sum_{ii}S^{T}Q_{ii}^{diag}.S+\sum_{(i<j)}S^{T}.Q_{ij}^{upper}.S
\end{eqnarray}
Consequently, the quadratic Ising function of spins can be written as below, 
\begin{equation}
E(J|h)=\sum_{ii}S^{T}Q_{ii}^{diag}.S+\sum_{(i<j)}S^{T}.Q_{ij}^{upper}.S \label{en}
\end{equation}
Here the vector $S=\{s_{i}\}$ with $s_{i}\in\{-1,+1\}$ can be mapped
to a vector $Y=\{y_{i}\}$with $y_{i}\in\{0,1\}$ by using a simple
arithmetic transformation $y_{i}=\frac{1+s_{i}}{2}$. This transformation is required to map the quadratic polynomial on available quantum annealer. After transformation
the condition $y_{i}^{2}=y_{i}$ is easily satisfied. The state $(0)$ represent
the spin down and the state $(1)$ represent spin up. Now the Eq.\ref{en} can be written 
in new variables $y_{i}$ as below,
\begin{eqnarray}
E(J|h)=\sum_{i}(2Y-1)^{T}.Q_{ii}.(2Y-1)\\+\sum_{i<j}(2Y-1)^{T}.Q_{ij}^{upper}.(2Y-1)
\end{eqnarray}
In short we can also write as,
\begin{equation}
E(J|h)=\sum_{i\leq j}(2Y-1)^{T}.Q_{ij}.(2Y-1)
\end{equation}
Point to be noted that, the diagonal terms of $Q_{ij}$ has the terms
$h_{ii}$ and upper triangular part of the matrix has the coupling
strengths $J_{ij}.$ So magnetic filed and coupling strength are absorbed
in the equation with variables $y_{i}\in\{0,1\}$. So we can also say
that the matrix $(2Y-1)$ encode the binary data with the variables
$y_{i}\in\{0,1\}$. So it is easy to re-express the equation explicitly
as below,
\begin{equation}
E(J|h)=\sum_{ii}X^{T}Q_{ii}X+\sum_{i<j}X^{T}Q_{ij}X
\end{equation}
Or,
\begin{equation}
E(J|h)=\sum_{i\leq j}X^{T}.Q_{ij}.X
\end{equation}
With the continuation of the above discussion, we can express the future possibilities of developing quantum algorithms by taking into consideration the adiabatic computation model. Apart from it the framework of quantum machine learning\cite{qml1,qml2} based on quantum annealing may also give new avenues for solving computational problems in HEP. 
\section{Conclusion}
The paper cover the perspective of quantum machine learning in high energy physics. There are many computational extensive tasks in high energy physics such track finding, particle identification, unfolding and many others. Specially track reconstruction is most time consuming task. We have described as the classical methods for track reconstruction and shown the possibility of usage of quantum annealing for the same in future. This introductory article deliver a lucid information towards the application of quantum computation in high energy physics. This may be helpful for experimental high energy physics community to produce the efficient quantum algorithms for computational problems in high energy physics.

\end{document}